\documentclass[aps,prl,twocolumn,superscriptaddress]{revtex4}
\usepackage{graphicx}

\begin{document}


\title{Charge Freezing in Zig-zag Chain Cuprates PrBa$_{2}$Cu$_{4}$O$_{8}$  
Observed by Cu Nuclear Quadrupole Resonance}

\author{S.~Fujiyama}
\altaffiliation{Present address: Institute for Molecular Science, 
Myodaiji, Okazaki, Aichi 444-8585, Japan}
\email{fujiyama@ims.ac.jp}
\author{M.~Takigawa}
\email{masashi@issp.u-tokyo.ac.jp}
\affiliation{Institute for Solid State Physics, University of Tokyo, 
Kashiwa, Chiba 277-8581, Japan}
\author{S.~Horii}
\affiliation{Department of Applied Chemistry, University of Tokyo, 
Bunkyo-ku, Tokyo 113-8586, Japan}

\date{\today}

\begin{abstract}
We report nuclear quadrupole resonance (NQR) studies on the chain Cu sites of 
PrBa$_{2}$Cu$_{4}$O$_{8}$, a quasi-one-dimensional conductor with a nearly 
quarter-filled band. The nuclear spin-lattice relaxation rate $1/T_{1}$ shows a 
pronounced peak near 100~K caused by fluctuations of electric field gradient (EFG).  
Similar peak was observed for the spin-echo decay 
rate $1/T_{2}$, however, at a different temperature near 50~K.  
These results and broadening of the NQR spectrum at low temperatures 
indicate that slow charge fluctuations of either electronic or ionic origin 
freeze gradually at low temperatures.  
\end{abstract}

\pacs{74.72.Jt, 76.60.Es, 76.60.Gv, 76.60.Lz} \maketitle

There has been increasing interest in quasi-one-dimensional 
(Q1D) correlated electrons. Theoretical studies on generalized Hubbard or 
$t$-$J$ models for chains and ladders have revealed rich phase diagram associated 
with various instabilities towards Mott localization, spin density wave, charge order, 
and superconductivity \cite{schulz941,giamar971, dagotto941}.  
A variety of phases has been indeed observed in organic Bechgaard salts 
\cite{gruner881} and cuprate ladder materials \cite{uehara961,nagata981}. 
Pr-based cuprates, PrBa$_{2}$Cu$_{3}$O$_{7}$ (Pr123) and 
PrBa$_{2}$Cu$_{4}$O$_{8}$ (Pr124), also provide good model systems of strongly 
correlated Q1D electrons. Each of these has identical structure 
to the respective Y-based high temperature superconductor, YBa$_{2}$Cu$_{3}$O$_{7}$ (Y123) and 
YBa$_{2}$Cu$_{4}$O$_{8}$ (Y124).  The CuO$_2$ planes in Pr-based compounds, 
however, show antiferromagntic order \cite{reyes901,fujiyama021} and 
their contribution to the optical conductivity revealed features 
of a charge-transfer insulator \cite{takenaka921, takenaka001}.  
The insulating nature of the CuO$_2$ planes was explained 
by Fehrenbacher and Rice \cite{fehren931} based on a model of the 
localized hole states made of Pr-4$f$ and O-2$p_{\pi}$ hybridized orbitals.  
Thus the active element for low energy spin and charge 
excitations in Pr-based cuprates are the Cu-O chains, single chains in Pr123 and 
double (zigzag) chains in Pr124. 

These two compounds show contrasting transport behavior.  
Pr124 is a highly anisotropic Q1D metal showing extremely large 
in-plane resistivity anisotropy $\rho_a/\rho_b \sim 1000$ at 4~K 
\cite{horii001,mcbrien021,hussey021}, where $b$ is along the chain 
direction and $a$ is perpendicular to it within the CuO$_2$ plane.   
Although $\rho_{b}$ shows monotonic metallic 
temperature dependence, $\rho_{a}$ exhibits a broad peak near 130~K, 
indicating incoherent transport perpendicular to 
the chains at high temperatures.  The angle resolved photoemission spectra (ARPES) 
\cite{mizokawa001} also revealed a clear one-dimensional dispersion which crosses 
the Fermi level near the momentum $k_{b}\simeq \pi/4$, pointing to a nearly 
quarter-filled conduction band.  In Pr123, the spectral weight of the 
chain contribution to the optical conductivity \cite{takenaka921} as well as the 
ARPES spectra \cite{mizokawa991} are also consistent 
with an approximately quarter-filled conduction band.  However, semiconducting 
behavior of the dc-resistivity, steep suppression of the optical 
conductivity below 0.1~eV and vanishing ARPES intensity near the Fermi 
level all indicate the existence of a charge gap.  Nuclear magnetic resonance (NMR) 
and nuclear quadrupole resonance (NQR) experiments in Pr123 \cite{grevin981} 
revealed line broadening and anomalies in relaxation rates, which were ascribed 
to a charge ordering instability.  

Apparent absence of static charge order in Pr124 may be due to 
slight deviation from the quarter-filling \cite{mizokawa001} or to the geometrical 
frustration of intersite Coulomb interactions in zigzag chains \cite{seo011}.  
However, dynamic signature towards charge order 
instability is indicated by the optical conductivity in Pr124, which 
splits into a zero-energy Drude part with a small weight and a finite energy mode 
centered at 40~meV \cite{takenaka001}.  In this letter, we report results of NQR 
experiments for the chain Cu sites in Pr124.  We found a pronounced peak in the 
temperature dependences of the nuclear 
spin-lattice relaxation rate ($1/T_1$) and spin-echo decay rate ($1/T_2$),  
as well as broadening of the NQR spectra, indicating gradual freezing of 
fluctuations of electric field gradient (EFG) with decreasing temperature.  

\begin{figure}[t]
\includegraphics*[width=8cm]{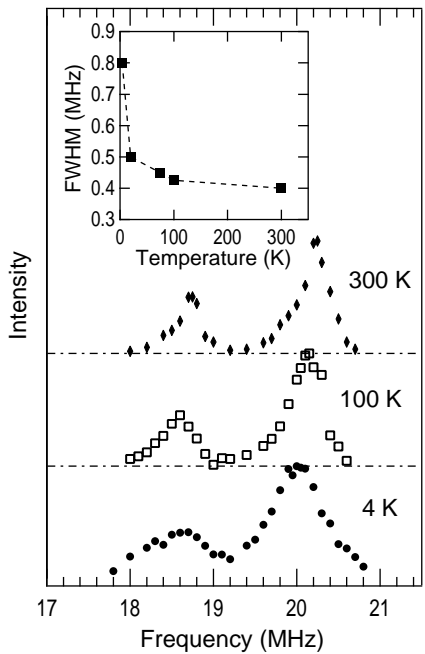}
\caption{\label{fig:spectra}NQR spectra for the chain Cu sites in Pr124. The peak near
20.2~MHz (18.7~MHz) is due to $^{63}$Cu ($^{65}$Cu) nuclei.  The inset shows
the temperature dependence of the line width (FWHM) for $^{63}$Cu.}
\end{figure}
The powder sample of Pr124 was synthesized by solid state reaction under high 
pressure as described in \cite{yamada941}.  Standard spin-echo pulse technique 
was used to obtain NMR spectra.  The spin-echo decay rate ($1/T_2$) 
was obtained from the spin-echo intensity measured against
the time separation $\tau$ between $\pi/2$ and $\pi$ pulses. 
The nuclear spin-lattice relaxation rate ($1/T_1$) was measured by inversion recovery 
method. 

The NQR spectra for the chain Cu sites in Pr124 are presented in 
Fig.~\ref{fig:spectra}. We observed well resolved resonance lines for both 
$^{63}$Cu and $^{65}$Cu isotopes with the full-width at half-maximum (FWHM) 
of 560~kHz for $^{63}$Cu at 300~K.     
The resonance signal from the planar Cu sites was observed near 70~MHz 
at 1.5~K as reported elsewhere \cite{fujiyama021}, indicating an 
antiferromagnetic order of the planer Cu spins.  
The NQR spectrum for the chain Cu sites shows little change at the 
N\'{e}el temperature of the planar Cu spins ($\sim 220$~K).  

\begin{figure}[t]
\includegraphics*[width=8cm]{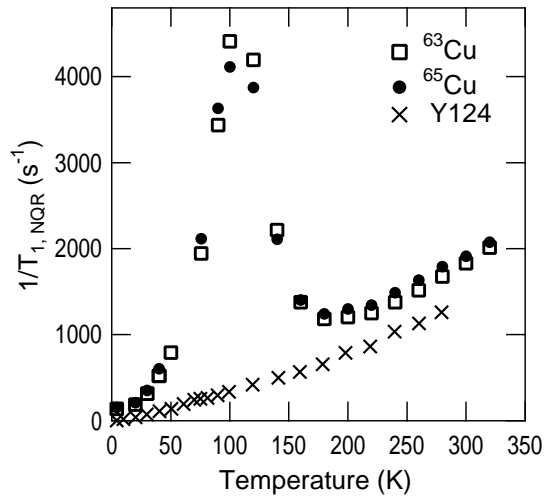}
\caption{\label{fig:T1}The temperature dependence of $1/T_{1}$ for $^{63}$Cu 
and $^{65}$Cu. The data for the chain $^{63}$Cu sites in Y124 
\cite{machi911}are shown for comparison.}
\end{figure}

The temperature dependence of $1/T_{1}$ for the chain Cu sites is shown in Fig.~\ref{fig:T1}.  
Here $1/T_1$ is defined as the inverse time constant for the exponential recovery 
of the NQR intensity after the inversion pulse,  
which is three times larger than the standard definition if the relaxation is due to 
magnetic processes.  
The fitting of the recovery curve to an exponential function was satisfactory over two 
decades except below 100~K, where the distribution of $1/T_1$ results in a slightly 
non-exponential recovery curve. 
Above 200~K, $1/T_1$ is nearly proportional to $T$, similar to the results
in Y124 \cite{machi911}.  
Below 200~K, however, $1/T_{1}$ deviates significantly from the $T$-linear behavior 
and shows a pronounced peak near 100~K.  
 
\begin{figure}[b]
\includegraphics*[width=8cm]{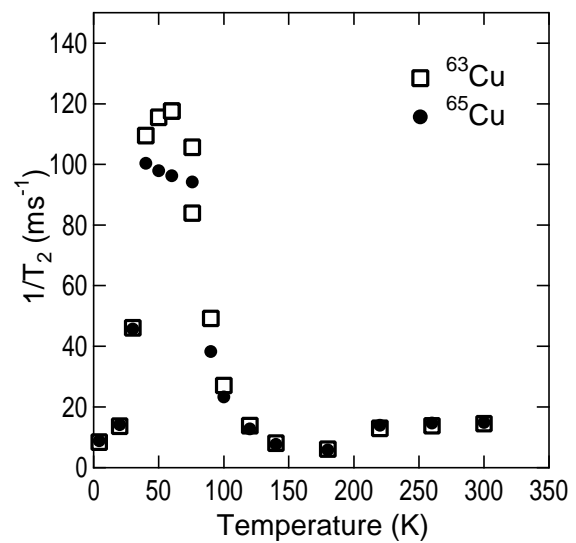}
\caption{\label{fig:T2}The temperature dependence of $1/T_{2}$ for $^{63}$Cu 
and $^{65}$Cu nuclei.}
\end{figure}

A similar peak was observed also for $1/T_{2}$ shown in Fig.~\ref{fig:T2}.
The spin-echo decay curves can be fit well to an exponential function 
$\exp(-2\tau/T_2)$ below 200~K.
At higher temperatures, the spin-echo decay curve contains also a small Gaussian 
component and the fitting becomes less satisfactory.    
Although $1/T_{2}$ is almost independent of temperature above 100~K, there is a 
strong peak near 50~K.  Note that this temperature is lower than the peak of $1/T_1$ by
50~K.  The peak value $1/T_{2} \simeq 120$~ms$^{-1}$ 
is more than 5 times larger than the value reported for Y124
\cite{itoh921}. 

The isotopic ratio of these relaxation rates can be used to identify the relaxation process.  
Both $1/T_1$ and $1/T_2$ are larger for $^{65}$Cu than for $^{63}$Cu when the temperature
is sufficiently higher or lower than the peak (Figs.~\ref{fig:T1} and \ref{fig:T2}).  Their ratio is close 
to the ratio of squared nuclear gyromagnetic ratio 
$(^{65}\gamma /^{63}\gamma)^{2}=1.148$, indicating that relaxation is  
caused by fluctuations of the local magnetic field acting on nuclei.  
In the temperature range near the peak, on the other hand, both relaxation rates are
larger for $^{63}$Cu, with the isotopic ratio close to the ratio of squared 
nuclear electric quadrupole moments $(^{65}Q /^{63}Q)^{2}=0.856$.  
Therefore, we conclude that the peaks in the 
relaxation rates are caused by fluctuations of the electric field gradient (EFG) tensor, 
$V_{\alpha\beta} = \partial^2 V/\partial x_{\alpha}\partial x_{\beta}$, 
where $V$ is the electrostatic potential at the nuclear position.

\begin{figure}[t]
\includegraphics*[width=8cm]{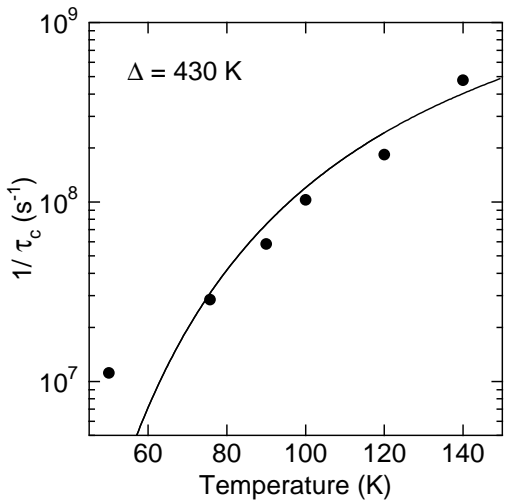}
\caption{\label{fig:tauc}The temperature dependence of the correlation time for 
EFG fluctuations. The line shows a fit yielding the activation energy of 430~K.}
\end{figure}

The relaxation rates can then be related to the time
correlation functions of EFG components.  A simple interpretation of 
temperature dependences of 
$1/T_{1}$ and $1/T_{2}$, in particular the difference in their peak   
temperatures, is given by a classical model of motional 
narrowing \cite{BPP,abragam}.  Let us assume a simple exponential form for the 
correlation functions, $\langle V_{\alpha\beta}(t)V_{\alpha\beta}(0) \rangle 
= \langle V_{\alpha\beta}^{2} \rangle \exp (-t/\tau_{c})$, where
$\tau_c$ is the correlation time of the fluctuations.  
The spin-lattice relaxation rate is given by their Fourier transform 
at the NQR frequency $\omega_n$ (20~MHz in our case),
\begin{equation}
\frac{1}{T_{1}}=\frac{\Delta^{2}\tau_{c}}{1+\omega_n^{2}\tau_{c}^{2}}, 
\label{eq:BPP} 
\end{equation}
where $\Delta$ is the magnitude of the transition matrix element of the 
quadrupolar interaction, which is proportional to the amplitude of the EFG fluctuations  
$\langle V_{\alpha\beta} \rangle$ appropriately averaged over different 
tensor components.   
If $\Delta$ is constant but $\tau_c$ depends on temperature,  
$1/T_1$ attains the maximum value when $\omega_n \tau_c=1$.
Thus the peak of $1/T_1$ is naturally understood due to increase of 
$\tau_c$ with decreasing temperature, i.e., slowing down of the EFG fluctuations
at low temperatures. 
By applying Eq.~(\ref{eq:BPP}) to the experimental data,
we deduced $\Delta/2\pi = 170$~kHz and $1/\tau_c$ is obtained as a function of
temperature near 100~K as shown in Fig.~\ref{fig:tauc}.  
Although the temperature range is limited (75~K $\geq T \geq$ 140~ K), we can extract an activation energy
430~K from the temperature dependence of $1/\tau_c$.   

In contrast, it is known that $1/T_2$ determined from the spin-echo decay curve
takes the maximum value $1/T_2 \sim \Delta$ when 
$\Delta \tau_c \sim 1$ \cite{takigawa861,abragam}.
Since $\Delta$ is two orders of magnitude smaller than $\omega_n$, gradual 
slowing down of the EFG fluctuations accounts for the fact the  
the peak in $1/T_2$ occurs at a lower temperature than
the peak of $1/T_1$.  The plot in Fig.~\ref{fig:tauc} indeed shows that 
$1/\tau_c \sim 10^6$~s$^{-1} \sim \Delta$ near 50K, where $1/T_2$ shows a peak. 
The value of $1/T_2$ at the peak (Fig.~\ref{fig:T2}) is also of the same orders 
of magnitude as $\Delta$, consistent with our model.   

If the correlation time continuously grows at lower temperatures, we expect the 
EFG fluctuations to freeze eventually resulting in broadening of the NQR spectrum. 
This is indeed observed as shown in Fig.~\ref{fig:spectra}. 
The width (FWHM) of $^{63}$Cu NQR spectrum is 400~kHz
at 300~K, while it increases rapidly below 20~K and reaches 800~kHz at 4.2~K.  
If we assume two independent contributions to the width, one due to
$T$-independent inhomogeneity which is represented by the spectrum 
at 300~K and the other due to freezing of EFG fluctuations at low temperatures, 
and square of the widths from these sources add to make up the observed 
value, the FWHM due to EFG freezing at 4.2~K is estimated to be 570~kHz.   
This value agrees with the second moment $\Delta/2\pi = 170$~kHz deduced from
the $1/T_1$ data, which translates to FWHM of 400~kHz
for a Gaussian distribution.  

The results of $1/T_1$, $1/T_2$, and the NQR spectra altogether 
provide convincing evidence for slow EFG fluctuations, which freeze randomly at 
low temperatures. Such fluctuations must be caused by motion of either electronic
or ionic charges.  Let us summarize the prominent features that characterize the 
fluctuations, irrespective of their origin. First, the dynamics 
is extremely slow, with $1/\tau_c$ ranging from $10^6$ to $10^9$ sec$^{-1}$ 
in the temperature range 50 - 150~K corresponding to the anomalous 
nuclear relaxation.  It is much too slow to be attributed to the transport of individual 
electrons along the chain.  Second is the glassy and random 
nature.  Gradual slowing down of the fluctuations without a well defined  
critical temperature is evidenced by the different peak temperatures 
of $1/T_1$ and $1/T_2$.  Spatial randomness with no signature of a 
long range order breaking translational symmetry is indicated by the 
broadening of the NQR spectrum with no line splitting nor fine structure.  
These are in sharp contrast to what have been observed in charge-density-wave (CDW)
materials such as NbSe$_3$ \cite{suits841}, where peak temperatures for 
$1/T_1$ and $1/T_2$ coincide.  Third, the amplitude of the fluctuating      
EFG is rather small, of the order of a few hundred kHz. It is known for high-$T_c$
cuprates that NQR frequency changes with hole concentration approximately at a rate
20 - 30 MHz/hole \cite{yasuoka951,zheng951}.  Therefore, if the NQR anomalies 
in Pr124 are due to electronic origin, only a minor fraction of the holes are relevant.
The charge distribution at low temperatures must still be largely uniform and only 
partial random freezing is possible with amplitude of the order of 1~\% of holes per site.  
 
Although we do not have conclusive evidence concerning the origin  
of the anomalous EFG fluctuations, following observations suggest that collective electron-lattice
coupled motion is important.  The peak in $1/T_1$ and the subsequent 
spectral change at lower temperatures described here are very similar to 
what were observed in the lightly hole-doped two-leg ladder compound 
Sr$_{24}$Cu$_{24}$O$_{41}$ \cite{takigawa981}.  This material is an insulator at low
temperatures and there are evidences for charge order in the ladder planes from
both the frequency dependent conductivity \cite{kitano011,blumberg021} and the
development of fine structure in the NQR spectrum \cite{takigawa981}.  Thus the peak in $1/T_1$ 
in this material is most likely caused by collective fluctuations of electronic charge.  
The similarity of the NQR results suggests that this is also the case for Pr124.
However, Pr124 is a good metal and dc-transport measurements indicate no sign of static charge order.  

The optical conductivity $\sigma(\omega)$ by Takenaka {\it et al} \cite{takenaka001}, 
on the other hand, clearly shows dynamical correlation towards charge
order instability, as mentioned earlier.  The spectrum of $\sigma(\omega)$ 
along the chain direction splits into the Drude like zero energy mode 
with only 2 \% of the total spectral weight and a gapped finite energy 
mode centered at 40 meV. The two-component structure resembles that 
observed in the organic Bechgaard salts \cite{dressel961,schwartz981} and suggests
dynamic short range correlation for charge disproportionation. The frequency dependence 
of high energy part of $\sigma(\omega)$ was analyzed in the framework of the
Tomonaga-Luttinger liquid, yielding a small value of the charge correlation 
exponent $K_{\rho} \sim 0.24$, which implies strong repulsive interaction \cite{takenaka001}.  
These results altogether point out that Pr124 is in close proximity to 
charge order, therefore, charge freezing may occur near impurities or can be 
triggered by a coupling to the lattice.  

No structural transition has been reported in Pr124 so far.  
However, recent neutron diffraction results \cite{back021} indicates 
sudden change of Ba position along the $c$-direction by about 0.05 $\AA$ near 160 K. 
The Raman spectra of Ba-phonon mode also show large frequency shift 
and sudden narrowing near the same temperature \cite{back001}.   
Although we notice a slight shift of the NQR frequency by about 100~kHz 
(Fig. \ref{fig:spectra}) from 300~K to 100~K, which may be a consequence of
movement of Ba, uniform change of ionic positions do
not cause broadening of the spectrum.  It is not likely that the 
static and dynamic NQR anomalies described here are merely lattice effects, 
since they are not observed in the isostructural Y124.  
On the other hand, when such double-well type ionic instability is coupled 
to an electron system close to charge order, we suspect that a static partial charge 
freezing can occur.

In summary, the pronounced peaks in $1/T_{1}$ and $1/T_{2}$ at 
different temperatures and broadening of the NQR spectrum at 
lower temperatures for the chain Cu sites in PrBa$_{2}$Cu$_{4}$O$_{8}$ provide 
evidence for slow fluctuations of electric field gradient.  
Their correlation time is extremely slow and 
show glassy freezing at low temperatures.  The amplitude of the fluctuations and 
concomitant freezing is about two orders of magnitude smaller than 
what we expect when the whole carrier in the Cu$_2$O$_2$ 
chains participate in charge order.   

\begin{acknowledgments}
We would like to thank H.~Seo, M.~Ogata and Y. Ohta for stimulating discussion and to M.~Osada
for information regarding Ref.~\cite{back021}. 
This work is supported by the Grants--in--aid for Scientific Research,
Priority Area (A) on the {\it Novel Quantum Phenomena in 
Transition Metal Oxides} from the Ministry of Education, Culture, Sports, 
Science and Technology Japan.  S. F. was supported
by the Research Fellowship for Young Scientists from the Japan Society 
for Promotion of Science.
\end{acknowledgments}


\end{document}